\documentclass[twocolumn,showpacs,preprintnumbers,amsmath,amssymb]{revtex4}

\usepackage{graphicx}
\usepackage{dcolumn}
\usepackage{bm}
\usepackage[dvips]{color}

\usepackage{float}
\usepackage{flafter}
\usepackage{epsfig}%

\newcommand{\gammadot}{\dot{\gamma}}

\newcommand{\sigmac}{\sigma^{+}_{\text{c}}}
\newcommand{\sigmacth}{\sigma^{+}_{\text{c,th}}}
\newcommand{\sigmacsim}{\sigma^{+}_{\text{c,sim}}}

\newcommand{\Tc}{T_{\text c}}

\newcommand{\NA}{N_{\text A}}

\newcommand{\tw}{t_{\text {w}}}

\newcommand{\rem}[1]{}
\newcommand{\com}[1]{}

\begin{document}

\preprint{Submitted to Phys. Rev. Lett.}
\title{Yield Stress Discontinuity in a Simple Glass}

\author{F. Varnik$^1$, O. Henrich$^2$}
\address{
$^1$Max-Planck Institut f\"ur Eisenforschung, Max-Planck Stra{\ss}e
1, 40237 D\"usseldorf, Germany\\
$^2$Fachbereich f\"ur Physik, Universit\"at Konstanz, 78457 Konstanz, Germany}

\date{\today}

\begin{abstract}
Large scale molecular dynamics simulations are performed to study
the steady state yielding dynamics of a well established simple
glass. In contrast to the supercooled state, where the shear stress,
$\sigma$, tends to zero at vanishing shear rate, $\gammadot$, a
stress plateau forms in the glass which extends over about two
decades in shear rate. This strongly suggests the existence of a
finite dynamic yield stress in the glass, $\sigma^+ (T) \equiv
\sigma(T; \gammadot \to 0)
>0$. Furthermore, the temperature dependence of $\sigma^+$ suggests
a yield stress discontinuity at the glass transition in agreement
with recent theoretical predictions. We scrutinize and support this
observation by testing explicitly for the assumptions (affine flow,
absence of flow induced ordering) inherent in the theory. Also, a
qualitative change of the flow curves enables us to bracket the
glass transition temperature $T_c$ of the theory from above and (for
the first time in simulations) {\it from below}. Furthermore, the
structural relaxation time in the steady state behaves quite similar
to the system viscosity at all studied shear rates and temperatures.
\end{abstract}
\maketitle

Soft glassy materials under shear exhibit a
rich phenomenology. In the dilute regime, at temperatures
corresponding to the liquid state, forced Rayleigh scattering
experiments \cite{Qiu} show an increase of diffusion
constant upon shearing (shear thinning), distinct from Taylor
dispersion (displacement of particles in the flow direction ($x$)
as they move in the direction of shear gradient. This would
give rise to $<\Delta x^2(t)>\sim t^3$!)
\cite{Taylor}. At higher densities, experiments show evidence for shear thinning
due to the presence of freely slipping two dimensional crystalline
layers  \cite{Ackerson}.

On the other hand, studies of disordered suspensions of hard spheres
show that shear thinning and shear melting phenomena may also occur
in the absence of a crystalline structure \cite{Petekidis}. Similar
observations have also been made in light scattering echo studies of
(disordered) dense emulsions
\cite{Hebraud}. Brownian dynamics
simulations show that shear thinning in concentrated colloidal
suspensions is related to the fact that, in the limit of low shear
rates, the main contribution to the shear stress originates from the
Brownian motion of colloidal particles and that this contribution
decreases with shear rate $\gammadot$ \cite{Phung}.

Recent theoretical progress, connecting nonlinear rheology with glass formation,
prompts us to study in detail by simulations the stationary states of a well established
glass model~\cite{Kob} under shear. We focus on the yielding behavior for constant shear rate
of glassy states close to but also far below the glass transition temperature.

Recently, Berthier, Barrat and Kurchan studied numerically a driven
spin glass and showed that shear thinning can be understood in terms
of an acceleration of inherently slow system dynamics by the
external drive \cite{Berthier}. Within this
approach, the stress depends on the shear rate via a power law (no
dynamical yield stress).  The 'soft glassy rheology model' of Sollich et al. \cite{Sollich}
 extends the minimal "trap model" originally
introduced by Bouchaud \cite{Bouchaud} in order to take into account
the effect of an external drive. The theory contains a noise
temperature, $x$, which controls the distance from the glass transition at $x=1$.
For $1<x<2$, a power law decrease of the stress with
applied shear rate is found, whereas in the jammed state ($x<1$),
a continuous onset of a dynamic yield stress is predicted, $\sigma^+ \equiv
\sigma(\gammadot \to 0)=1-x$.
Fuchs and Cates \cite{Fuchs}, on the other hand, started from the
well studied mode coupling theory (MCT) of the glass transition \cite{Goetze}. \
They started from the idealized
picture that, in a supercooled liquid, nearest neighbors of a
particle form a cage which progressively solidifies eventually
leading to a complete arrest of all particles as the glass
transition is reached. The effect of shear then enters by the
advection of density fluctuations.  Fluctuations of a given length scale are
advected towards progressively shorter length scales so that particles need explore
smaller regions in order for density correlations to decay.
The interesting prediction of a yield stress discontinuity at the ideal
glass transition was made. A related MCT approach to the fluctuations around the steady state
has recently been proposed by Miyazaki and Reichman \cite{Miyazaki}.
The issue of yield stress discontinuity, however, could not be
addressed in that approach.

Berthier and Barrat \cite{Berthier} performed
molecular dynamics simulation studies of the present model under a
homogeneous shear showing e.g.\ that, in a range of low shear rates,
time-shear superposition and space time factorization theorems hold
thus suggesting that generic properties related to the glass
transition "generalize" to the non-equilibrium situation of a
homogeneous shear. However, due to a rather limited range of shear
rates, results presented in Ref.
\cite{Berthier} did not allow a clear answer
whether the present model exhibits a yield stress or not.

In this paper, we focus exactly on this aspect, namely an analysis
of the dynamic yield stress and its behavior at the ideal glass
transition. For this purpose, we performed large scale molecular
dynamics simulations of a generic glass forming system first
introduced by Kob and Andersen \cite{Kob}. The model
consists of a 80:20 binary mixture of Lennard-Jones particles (whose
types we call A and B) at a constant total density. A and B particles interact
via a Lennard-Jones potential. Standard (dimensionless) parameters
will be used as in Refs.~\cite{Berthier,Kob}.

Ten independent samples, equilibrated at a temperature of $T=0.45$,
serve as starting configurations for all simulations reported here. 
The temperature is set to the desired value at the beginning of shear 
motion.  The shear stress is computed using particle positions and 
velocities via the Irving-Kirkwood formula \cite{Evans}. Only samples 
corresponding to the steady state, i.e.\ to strains larger than $100\%$, are taken into
account (previous studies of the stress-strain relation of the same
model showed that the initial transient behavior is limited to
strains below 50\% \cite{Varnik}). Depending on the desired accuracy
of the stress-data, the length of simulations was varied between
380\% and 760\% strain. Equations of motion are integrated using a 
discrete time step of $dt=0.005$.

\rem{Note that, using this time step and for a system of 1000
particles, a single run at the lowest simulated shear rate of $3\times
10^{-6}$ takes between two and four weeks on a 3GHz
AMD-Athlon CPU. However, in order to check for the accuracy of
results, we also performed simulations with a smaller time step of
$dt=0.002$ at a temperature of $T=0.42$ (close to $\Tc$) and for
shear rates ranging from $\gammadot=10^{-1}$ to $\gammadot=10^{-5}$.
All results obtained with this time step were identical to those
with $dt=0.005$.}

Recently, it was found that the present model may exhibit
shear-localization in the glassy state if the shear rate is imposed
by using a conventional Couette cell with moving atomistic
walls~\cite{Varnik}. In the present analysis, however, we are
interested in effects of a spatially constant shear rate, a basic
ingredient of all theories briefly addressed above. Therefore, we do
not use atomistic walls but apply the so called SLLOD algorithm
combined with Lees-Edwards boundary condition \cite{Evans}. With
this simulation method, we do indeed observe a linear velocity
profile in all studied cases (not shown).

\begin{figure}
\unitlength=1mm
\begin{picture}(160,41)
\put(5,-4){
\epsfig{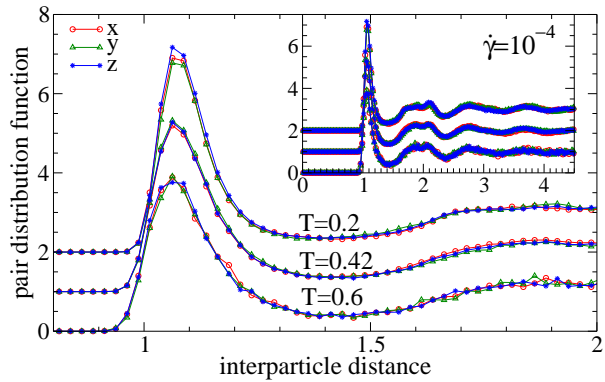}} 
\end{picture}
\caption[]{A-A pair correlation functions along $x$ (flow),
$y$ and $z$ (shear gradient) directions, for three characteristic
temperatures: $T=0.2$ (glass), $T=0.42$ (close to $\Tc$) and $T=0.6$
(supercooled state). The shear rate is $\gammadot=10^{-4}$. For
clarity, data at $T=0.42$ ($T=0.2$) are shifted upwards by 1 (2).}
\label{fig:gxyz}
\end{figure}

As shear may, at least in principle, change the static structure of
the system, we examine this by computing pair distribution functions
along $x$ (flow), $y$ and $z$ (shear gradient) directions
separately. \rem{For the $x$ direction, for example, we compute the
(majority) A-A pair correlation via $g(x)=\langle\sum_{i=1}^{\NA}
\sum_{j>i}
\delta(|x_i-x_j|-x)\delta(y_i-y_j)\delta(z_i-z_j)\rangle$, and
finally normalize the result by the ideal gas value so that, in the
absence of long range order, $g(x\to \infty)=1$ is expected ($\NA$
is the number of A-particles).} Figure \ref{fig:gxyz} illustrates
results on $g(x)$, $g(y)$ and $g(z)$ (A-A correlations) for three
characteristic temperatures $T=0.2$ (deep in the glass), $T=0.42$
(close to $\Tc$) and $T=0.6$ (supercooled state) indicating the
absence of long range order in all studied cases. Surprisingly, even
though the stress changes by more than a decade at this $\gammadot$,
the local structure varies little with shear, remains amorphous, and
(almost) isotropic. Similar observations are also made using the B-B
and A-B pair correlations.

\begin{figure}
\unitlength=1mm
\begin{picture}(160,50)
\put(1,-2){
\epsfig{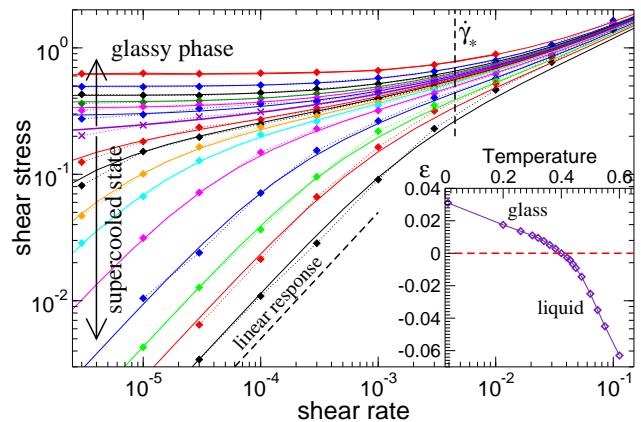}}
\end{picture}
\caption[] {Simulated shear stress (symbols)
compared to theoretical predictions (solid lines) for various
temperatures ($T=0.01$ to $T=0.60$ from top to bottom). The inset shows 
the fitted separation parameter, $\epsilon$, versus $T$.} \label{fig:stress}
\end{figure}

Figure \ref{fig:stress} shows simulated steady state shear stresses
as functions of shear rate (viz. 'flow curves') for temperatures ranging from
far above to far below the glass transition temperature of the model.
Note that shearing allows us to access stationary states at very low
temperatures. As apparent from the change of the curvature of the
flow curves (S--shaped without extended horizontal piece at high,
horizontal piece merging into upward curvature at low temperatures)
the system response changes qualitatively around a temperature
$\Tc$, which MCT identifies as (ideal) glass transition temperature.
For $T>\Tc$ the stress becomes proportional to shear rate as
$\gammadot$ approaches zero (linear response). At temperatures below
$\Tc$, however, a stress plateau forms in the low
$\gammadot$-regime. The observed qualitative agreement with the MCT scenario 
prompts us to test its predictions in  more detail.
Importantly, the qualitative change manifest in
the $\sigma(\gammadot)$ curves enables us to give upper and lower
limits for $\Tc$ without any theoretical analysis: we conclude $0.34
< \Tc < 0.45$. Fits with the schematic $F^{\gammadot}_{12}$-model of
MCT support this estimate, and achieve to describe the
flow curves for \emph{all} studied
temperatures by adjusting two global parameters (a scale for
$\gammadot$ and one for $\sigma$), and the parameter $\varepsilon$ at each
temperature measuring the distance to the ideal glass transition
\cite{Henrich}. The fit gives $\Tc = 0.4$ and $\sigmacth=0.19$.
We thus conclude that the basic rheological features of our model are
well described within simple schematic models in the framework of the idealized MCT.
Relaxation channels not contained in the idealized MCT (so--called
'hopping effects' \cite{Kob}, which may be the origin for the
deviations from theory at very low $\gammadot$ in Fig.\
\ref{fig:stress}) can not falsify our bound for $\Tc$, because the
qualitatively different shapes of the flow curves are the
characteristics of the fluid or glassy states within MCT.

\begin{figure}
\unitlength=1mm
\begin{picture}(160,42)
\put(5,-4){
\epsfig{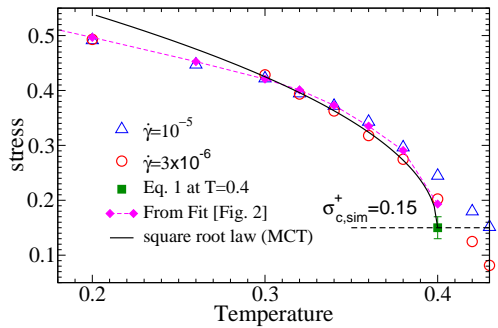}}
\end{picture}
\caption[]{Determination of dynamic yield stress and its temperature
dependence (see text).} \label{fig:sigma+}
\end{figure}

The stress plateau is best developed for temperatures deep
in the glassy phase extending over about two decades in shear rate.
Its onset is shifted toward progressively
lower $\gammadot$ as the temperature is increased toward $\Tc$.
This makes an estimate of the dynamic yield stress, $\sigma^+(T)
\equiv \sigma(T; \gammadot \to 0)$, a difficult task for
temperatures below but close to $\Tc$.
Nevertheless, an estimate of $\sigma^+(T)$ is interesting because it
highlights the anomalous weakening of the glass when heating to $T_c$.
Testing the MCT predictions below $T_c$ has previously not been possible in
simulations because of problems to reach the
equilibrated or steady state at sufficiently low
shear rates. Our estimate is obtained by comparing the steady state
shear stress for the two lowest simulated shear rates, namely
$\gammadot=10^{-5}$ and $\gammadot=3 \times 10^{-6}$. As shown in
Fig.\ \ref{fig:sigma+}, at temperatures below $T=0.38$, practically
the same shear stress is obtained for both choices of $\gammadot$
indicating the presence of a yield stress plateau.

For $T=\Tc$, we make use of theoretical predictions based on the
$F^{\gammadot}_{12}$-model \cite{Henrich}. For not too low shear rates,
the flow curve takes the form of a generalized Hershel-Bulkeley constitutive
equation,
\begin{equation}
\sigma=\sigma^+_{\text{c}} \left( 1+|\gammadot/\gammadot_*|^m +c_2
|\gammadot/\gammadot_*|^{2m} + c_3 |\gammadot/\gammadot_*|^{3m}
\right). \label{eq:sigma}
\end{equation}
Here, $\gammadot_*$ is an upper limit where this expansion holds. As
the simple $F^{\gammadot}_{12}$-model is found to describe the
rheological properties of our system rather well (see the discussion
of Fig.\ \ref{fig:stress}), we set the parameters, $c_2=0.896$,
$c_3=0.95$ and $m=0.143$ as obtained from an analysis of the model
\cite{Henrich}. We then apply a fit to Eq.\ \ref{eq:sigma} with
$\sigmac$ and $\gammadot_*$ being the only fit parameters. This
gives  $\sigmacsim=0.15 \pm 0.01$ and $\gammadot_*=0.0045 \pm 0.0008$
the latter being close to our estimate of the window, where
MCT can describe the flow curves. At higher shear rates,
$\gammadot>\gammadot_*$, we expect microscopic effects to dominate
the stress. As shown in Fig.\ \ref{fig:sigma+}, while $\sigma^+(T)$
weakly varies with $T$ at low temperatures, it steeply drops as $T$
approaches $\Tc$, signalling the glass transition. The yield stress
follows well the MCT-square root law \cite{Fuchs},
$\sigma^+(T)-\sigma^+_{\rm{c}} \propto |1-T/\Tc|^{0.5}$.

Note that the critical temperature of $\Tc=0.4$, used in order to
obtain best agreement between the theory and simulations, is
slightly lower than the estimate $\Tc=0.435$ obtained from the
analysis of the equilibrium dynamics of the system \cite{Kob}.  This
discrepancy is possibly related to the hopping effects observed in
the density correlation functions closely above $\Tc$
\cite{Kob,Flenner}. A closer analysis of this aspect requires
understanding of hopping effects under shear and is beyond the scope
of the present report.

\begin{figure}
\unitlength=1mm
\begin{picture}(160,40)
\put(5,-4){
\epsfig{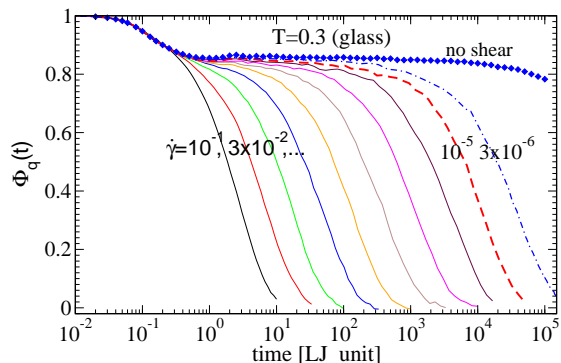}}
\end{picture}
\caption[]{Steady state incoherent scattering function,
$\Phi_q(t)$, of a glass ($T=0.3$)  measured at various shear rates as indicated,
and for $q=7.1$. The quiescent $\Phi_q(\gammadot=0)$ after a waiting
time of $\tw=10^5$  is also shown.} \label{fig:phiq}
\end{figure}

\begin{figure}
\unitlength=1mm
\begin{picture}(160,40)
\put(5,-4){
\epsfig{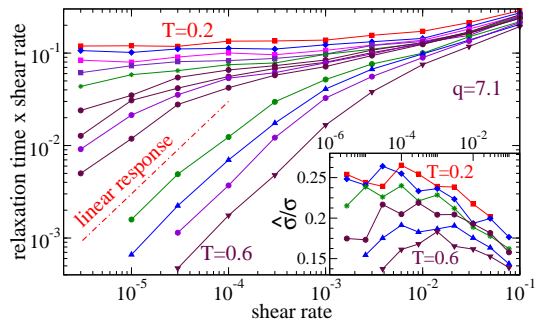}}
\end{picture} 
\caption[]{$\hat{\sigma} \equiv \tau_q \gammadot$ versus
shear rate for various temperatures ranging from the glassy phase to
the supercooled state. The inset shows the ratio of this (Maxwell
type) stress to the real shear stress for $T=0.2,\; 0.3,\; 0.4, \;
0.45,\; 0.525$ and $0.6$.} \label{fig:taugdot}
\end{figure}

Next we address the close connection between observed non-linearity
in the flow curves and the structural relaxation as reflected in the
incoherent scattering function, $\Phi_q(t) = < \sum^{\NA}_{i=1}
\exp[q_y(y_i(t)-y_i(0))] >/\NA$. Here, $\NA$ is the number of
A-particles, $q_y$ is the $y$-component of the wave vector and $y_i$
is the $y$-component of the position of $i$-th particle. We set
$q_y=7.1$ to the inverse of the average interparticle distance (or,
equivalently, the  maximum position of the static structure factor).
The above definition of $\Phi_q(t)$ takes into account the fact that
the imposed flow is in the $x$-direction and the shear gradient in
the $z$-direction, and eliminates the advection of particles with
the flow. Note that the above choice of A type is arbitrary. Similar
results are also obtained for type B particles (not shown).

Figure \ref{fig:phiq} illustrates $\Phi_q(t)$ for $T=0.3$ (glass)
for all studied shear rates. First, $\Phi_q(t)$ exhibits the two
step relaxation process typical of supercooled liquids: A short time
decay to a plateau, $f_q$ (characterizing the 'solidity' of the
system at length scale $1/q$ \cite{Goetze}) followed by a final
decay to zero at much larger times. The long time decay of
$\Phi_q(t)$, on the other hand, is clearly dominated by the imposed
shear. Its shape does not depend on the shear rate which only sets
the time scale, $\tau_q \sim1/\gammadot$. Note that Fig.\
\ref{fig:phiq}, and the stationarity of the correlators, verifies
that we achieved ergodic stationary states as expected from theory.

Figure \ref{fig:taugdot} depicts $\hat{\sigma}=\tau_q \gammadot$
(this definition of a stress goes back to Maxwell) for $q=7.1$
versus $\gammadot$ for various temperatures. Here,
$\tau_q=\int_0^{\infty} dt \Phi_q(t)$ is defined as average
relaxation time. Where $\tau_q \sim1 / \gammadot$ holds, a plateau
in $\hat\sigma$ follows. Note the striking similarity between the
$\gammadot$-dependence of the shear stress ($\sigma$; Fig.\
\ref{fig:stress}) and that of $\hat{\sigma}$. As suggested by the
inset of Fig.\ \ref{fig:taugdot}, in the whole range of studied
shear rates and temperatures, $\hat{\sigma} / \sigma$ changes at
most by a factor of two whereas the shear stress varies by two
orders of magnitude. Thus, both the $\gammadot$ and $T$-dependence
of steady state shear stress is mainly determined by that of
$\hat{\sigma}=\gammadot\tau_q$. This observation is quite
significant as it supports theoretical approaches where, even beyond
the Newtonian--regime, the shear viscosity  is simply taken as a
relaxation time \cite{Berthier,Miyazaki}, or where this relation
holds as approximation \cite{Fuchs}.

In summary, large scale molecular dynamics simulations have been
performed in order to investigate the existence and temperature
dependence of the dynamic yield stress, $\sigma^+$, for a 80:20
binary Lennard-Jones model glass first proposed by Kob and Andersen
\cite{Kob}. Our data do indeed support the existence of a dynamic
yield stress in the glassy phase as underlined by stress plateaus
extending over about two decades in shear rate.
Let us mention recent experiments on the rheology of dense colloidal dispersions
\cite{Fuchs2005} which also find finite $\sigma^+$.
Furthermore, the temperature dependence of $\sigma^+$ follows the predicted
anomalous weakening close to the glass transition with a finite critical yield stress
of $\sigmac=0.17 \pm 0.02$ as a compromise between $\sigmacth=0.19$ and
$\sigmacsim=0.15$. The flow curves allow for the first bracketing in simulations of
the critical temperature of MCT  {\em from below}. Irrespective of
hopping effects neglected in the employed MCT, we can conclude $\Tc>0.34$.

Furthermore, a generalized stress ($\hat{\sigma}$; Fig.\
\ref{fig:taugdot}) as the product of the shear rate and the
structural relaxation time was determined for all temperatures and
shear rates. $\hat{\sigma}$ exhibits exactly the same qualitative
features as the real shear stress thus emphasizing the close
connection between the structural relaxation and the rheological
response. Noting that a priori no relation $\hat{\sigma}$ to
$\sigma$ exists \cite{Fuchs}, this finding
corroborates theoretical approaches which make use of the relaxation
time as a sort of viscosity.

{\noindent \bf Acknowledgments}\\
We thank J.-L.\ Barrat, L.\ Berthier, L.\ Bocquet, M.~E. Cates and J.\ Horbach for useful
discussions. F.V. was supported by the DFG, Grant No VA 205/1-1, and O.H. by the DFG
IGC 'Soft Matter'. Simulation time was granted by the ZDV-Mainz and PSMN-Lyon and IDRIS
(project No 031668-CP: 9).

\end{document}